\documentclass[aps,pra]{revtex4}   

\usepackage{amsmath}    
\usepackage{graphicx}   
\newcommand{\ket}[1]{\left|#1\right\rangle}
\newcommand{\0}{\ket{0}}
\newcommand{\1}{\ket{1}}

\begin{document}

\title{A counterfactual Rydberg gate for photons}
\author{Juan Carlos Garcia-Escartin}

\affiliation{Universidad de Valladolid, Dpto. Teor\'ia de la Se\~{n}al e Ing. Telem\'atica, Paseo Bel\'en n$^o$ 15, 47011 Valladolid, Spain}
 \email{juagar@tel.uva.es}   
\author{Pedro Chamorro-Posada}
\affiliation{Universidad de Valladolid, Dpto. Teor\'ia de la Se\~{n}al e Ing. Telem\'atica, Paseo Bel\'en n$^o$ 15, 47011 Valladolid, Spain}
\date{\today}

\begin{abstract}
Quantum computation with photons requires efficient two photon gates. We put forward a two photon entangling gate which uses an intermediate atomic system. The system includes a single Rydberg atom which can switch on and off photon absorption in an ensemble using the dipole blockade. The gate is based in a counterfactual protocol. The mere possibility of an absorption that can only occur with a vanishing probability steers the photons to the desired final state.
\end{abstract}
\maketitle

\section{Quantum computers and their implementations}
\label{intro}
Quantum computers would be able to run new protocols and algorithms, some of them more efficient than any classical alternative \cite{NC00}. There are several requisites for the practical implementation of a quantum computer: the data must be easy to prepare and read, the computer has to be able to perform any quantum logical operation and maintain coherence during the processing and the system must be scalable \cite{DiV00}. 

There are many candidate systems, each with its own strong and weak points. An interesting family of proposals uses photons as information carriers \cite{Ral06}. Photons have long coherence times and good transmission properties, can be produced and detected with reasonable efficiency and the state of a single photon can be manipulated with standard linear optics equipment \cite{KMN07}. The main obstacle to quantum computation with photons is the creation of two photon gates. 

Controlled operations in which the state of one photon alters the state of a second photon are essential for universal quantum computation \cite{DiV95}. Two photon interaction is elusive. Most two photon gate proposals either require strongly nonlinear media with losses that can make the operation inefficient \cite{Mil89}, need a number of resources that grows exponentially with the size of the system \cite{CAK98}, or use measurement induced nonlinearities \cite{PFJ03,OPW03,GPW04} that introduce a probabilistic element (the gates only work correctly with a certain low, theoretically bounded, probability \cite{Kni02,Kni03}). Proposed alternatives also include the use of weak nonlinearities \cite{MNS05} or measurement assisted probabilistic gates \cite{KLM01}.

We propose a controlled sign gate in which the photon interaction is mediated by coupled Rydberg atoms. The proposal tries to make the atomic part as simple as possible. The gate is based on a version of the quantum Zeno effect, in which frequent measurement steers the evolution of a quantum state \cite{MS77,IHB90}.

\section{Optical encoding: dual-rail}
\label{encoding}
\begin{figure}[ht]
\centering
\includegraphics[scale=0.95]{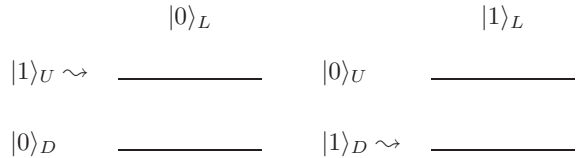}
\caption{Dual-rail path encoding. The logical $\0_L$ state is represented with a photon taking the upper path. In the $\1_L$ state, the photon travels through the lower path. \label{dualrail}} 
\end{figure}

We consider a dual-rail encoding representation. The basic logical quantum states $\0_L$ and $\1_L$ are encoded with a single photon in different orthogonal modes. One usual encoding uses orthogonal polarizations, like horizontally and vertically polarized photons so that $\0_L\equiv\ket{H}$ and $\1_L\equiv\ket{V}$. We will use an alternative encoding with spatial separation. We define two paths, each containing a number state $\ket{n}$, where we can have $n=0$ (no photons) or $n=1$ (a single photon). We label the paths as up and down (with subindices $U$ and $D$). There will only be one photon. The logical zero state is represented by a photon in the upper mode $\0_L\equiv\1_U\0_D$ and the logical one state by the photon taking the lower path $\1_L\equiv\0_U\1_D$ (see Figure \ref{dualrail}). Any superposition of logical values is encoded as a photon in a superposition of different trajectories. 

\section{CNOT and CZ gates}
\label{CNOTCZ}
In dual-rail encoding, any single photon logical operation can be implemented with beamsplitters and phase shifters alone \cite{RZBB94}. In order to produce any desired quantum operation, we only need a CNOT gate or, equivalently, a CZ gate \cite{BBC95}. The two interacting photons are called the control and the target photon. The state of the target photon is changed depending on the logical value of the control photon.

For a CNOT gate, if the control state is $\1_L$, the target photon changes its value (from $\0_L$ to $\1_L$ or from $\1_L$ to $\0_L$). In our encoding, if the control photon is in mode $U$, there is no change in the target photon. If it is in the lower path, the up and down paths of the target photon are switched.

A CZ gate introduces a $\pi$ phase shift between the upper and the lower paths of the target photon whenever the control photon is down. 

These two gates are essentially equivalent. We can convert a CZ gate into a CNOT, and vice versa. We only need two Hadamard gates. In our encoding, they can be implemented with a 50\% beamsplitter where the inputs are the upper and lower paths of the target photon. The beamsplitter divides the photon state into an equal superposition of both paths. If there are no sign changes, at the second beamsplitter there is a constructive interference in the original up or down port, while there is a destructive interference in the originally empty path. A sign shift inverts the ports of the constructive and destructive interferences at the second beamsplitter, changing the path of the photon. 

\section{Quantum interrogation}
\label{QI}
Quantum interrogation is a counterfactual protocol in which an absorbing object can be detected with a single photon that is not absorbed during the process. We discuss a scheme based on the Elitzur-Vaidman bomb test \cite{EV93}. We imagine a bomb so sensitive that even a single photon is enough to make it explode. We would like to detect whether there is such a bomb or not in a certain region of space without triggering its explosion. With a quantum interrogation setup we can detect the bomb using a single photon which is not absorbed \cite{KWH95,KWM99}. 

We assume that the bomb explosion is a macroscopic event equivalent to a quantum measurement. The whole protocol is based on the effects of negative measurement in quantum mechanics: not finding a photon in one path alters the state of the photon, even though the photon and the detector cannot have interacted in a classical sense. 

\begin{figure}[ht]
\centering
\includegraphics{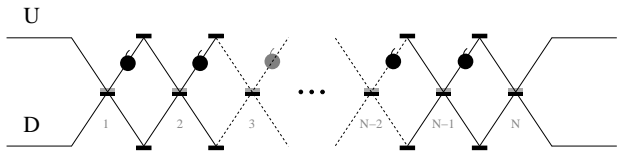}
\caption{Quantum interrogation: Looking at the path of a single photon, we can detect a bomb without any classical interaction.\label{QIpath}} 
\end{figure}

Take the setup of Figure \ref{QIpath} with two photon paths, up and down, and an input state $\0_U\1_D$, with one photon in the lower port. We would like to know if there is a bomb in the upper half of the system. We consider a large bomb occupying all the upper path (represented as a series of bombs in the photon's trajectory). The path of the photon is directed using mirrors (the black rectangles) and beamsplitters of reflectivity $\cos^2 \theta$ and transmittance $\sin^2\theta$. The beamsplitters are represented with rectangles with a grey side. Transmission of a photon coming from the grey side introduces a $\pi$ phase shift. 

Take the evolution when there is a bomb. The first beamsplitter turns the input photon state $\0_U\1_D$ into the path superposition $\cos\theta\0_U\1_D+\sin\theta\1_U\0_D$. The bomb explodes with probability $\sin^2\theta$ (the probability of the measurement finding the photon up). With a probability $\cos^2 \theta$ the bomb does not explode and the state of the photon goes back to $\0_U\1_D$. At the next beamsplitter, this evolution is repeated. After the $N$ beamsplitter stages, the state is still $\0_U\1_D$ with a probability $\cos^{2N} \theta$. For a small $\theta$, the probability of success can be approximated by $\left(1-\frac{\theta^2}{2}\right)^{2N} \approx 1- N\theta^2$, which can be made arbitrarily close to 1 by choosing $N\theta^2\ll 1$. 

If there is no bomb, each beamsplitter introduces a $\theta$ rotation. The state after the first beamsplitter is $\cos\theta\0_U\1_D+\sin\theta\1_U\0_D$, after the second $\cos(2\theta)\0_U\1_D+\sin(2\theta)\1_U\0_D$, and so on. At the output of the last beamsplitter the photon state is $\cos(N\theta)\0_U\1_D+\sin(N\theta)\1_U\0_D$.

In the usual setup, we fix $\theta=\frac{\pi}{2N}$ so that the final state is $\0_U\1_D$ if there is a bomb and $\1_U\0_D$ if there is none. We can measure both paths. A photon found in the lower port means there is a bomb, a photon up guarantees its absence. For high $N$, the bomb is detected without explosion with high probability ($1-\frac{\pi^2}{4N} \to 1$). 

The frequent measurement induced by the bomb freezes the state of the photon. The bomb and the photon cannot have met in any classical sense without triggering the explosion. However, the possibility of interaction allows a method for detection. 

In our two photon gates we use an alternative configuration with $\theta=\frac{\pi}{N}$. At the output we have state $\0_U\1_D$ if there is a bomb and a sign shift (state $-\0_U\1_D$) if there is no bomb. The probability of explosion is only $\frac{\pi^2}{N}$, which vanishes for a large $N$. The sign shift is not relevant for a single photon, but we will use a similar setup in which we can have a relative phase shift between photons on certain situations. This gives a CZ operation.

\section{Quantum interrogation gates}
\label{QIgates}
If the bomb is quantum, quantum interrogation correlates the state of the photon and the bomb \cite{Hor01,ZZG01}. Imagine a bomb which can be in a superposition of being in the path of the photon (state $\0_B$) and out of the photon's way (state $\1_B$). The bomb can be in any superposition $\alpha\0_B+\beta\1_B$ of these two states, with $\alpha$ and $\beta$ complex and $|\alpha|^2+|\beta|^2=1$. For an input photon in state $\0_U\1_D$ and $\theta=\frac{\pi}{2N}$, we have the evolution
\begin{equation}
\left(\alpha\0_B+\beta\1_B\right) \0_U\1_D \longrightarrow  \alpha\0_B\0_U\1_D +\beta\1_B\1_U\0_D. 
\end{equation}
The final position of the photon becomes associated to the position of the bomb.

This kind of scheme has been proposed as a way to generate entangled photon states \cite{GWM02,Azu03,Azu04} and can also be used to give a CNOT gate between two atoms inside separate cavities \cite{MW01}. We will discuss a simplified version of our quantum interrogation CNOT gate between light and matter with two coupled polarization interferometers \cite{GC06b}. We imagine the bomb is an atom, condensate or some other matter system which can absorb or scatter a photon and be in a superposition of being absorptive and non-absorptive. There have been further proposals for quantum interrogation gates and entanglement generators with different kinds of systems which are, essentially, based on the same principle we are going to discuss \cite{Pav07,WYN08,HM08,GC08b,GC09,SWC09}. 

Apart from these implementations, we can have a different family of Zeno gates in which there are two photons, each playing the role of the bomb for the other (if they meet, they destroy each other) \cite{FJP04,LR06,MG07,FPJ07,BOS11}. Conceptually, this is similar to the Hardy paradox with two antiparticles \cite{Har92}. Here, the interaction happens in a two photon absorption medium which absorbs pairs of photons but lets single photons pass.

Our proposal tries to relax the requisites on the quantum bomb system. We put forward a system based on Rydberg atoms which only requires a limited control on the atomic part and has a photon-bomb coupling mechanism which does not imply high quality resonant cavities. Instead of a CNOT gate, we propose a scheme for the computationally equivalent CZ gate. We take a single photon in a superposition of the $\0_L$ and $\1_L$ logical states. The lower photon state $\ket{n}_D$ will be directed to the lower path of the quantum interrogation setup of Figure \ref{QIpath}. The upper photon path is taken apart from the quantum interrogation interferometer in a detour. The length of the path of the photons in and out of the interferometer is adjusted to be equal to avoid unintended phase shifts between the logical $\0_L$ and $\1_L$ states. We suppose we can have any superposition of both photon trajectories, $\alpha\0_L+\beta\1_L\equiv \alpha\1_U\0_D+\beta\0_U\1_D$. 

If $\theta=\frac{\pi}{N}$, at the output we have
\begin{equation}
\label{EqCZ1}
\0_B(\alpha\1_U\0_D+\beta\0_U\1_D)\longrightarrow \0_B(\alpha\1_U\0_D+\beta\0_U\1_D)
\end{equation}
and
\begin{equation}
\label{EqCZ2}
\1_B(\alpha\1_U\0_D+\beta\0_U\1_D)\longrightarrow \1_B(\alpha\1_U\0_D-\beta\0_U\1_D).
\end{equation}

When the photon is up, it does not enter the quantum interrogation setup and the state doesn't change for any bomb state. If the photon enters the setup and there is a bomb, the negative measurement it provides freezes the photon to its original state. The sign shift only happens when the photon is down and there is no bomb. The most relevant part is the relative phase shift between the $\1_U\0_D$ and the $\0_U\1_D$ terms when the bomb state is $\1_B$, which gives the CZ evolution. The existence of an upper path (of the same length) takes a trivial transformation and makes it significant. We discuss an implementation of the quantum bomb with Rydberg atoms.

\section{Quantum bombs with Rydberg atoms}
\label{RydbergBombs}
A good quantum bomb must be able to be in a superposition of being perfectly absorptive and perfectly transparent. We suggest using an atomic ensemble with Rydberg atoms that show dipole blockade. The given scheme could be implemented in a cold Rydberg gas, for which there are effective manipulation techniques \cite{LFC01}. Alkali vapour cells, with rubidium or cesium, are a suitable support. 

We are going to work with ground level $\ket{g}$ and the long-lived Rydberg state $\ket{r}$ of a single Rydberg atom. There are many states from the ground state manifold and from the Rydberg excited states which could be used. In our quantum bomb there is a single Rydberg atom and an ensemble of $N_a$ Rydberg atoms in a collective state. For the ensemble, we define a state $\ket{\bf{g}}=\ket{g_1}\ket{g_2}\cdots\ket{g_{N_a}}$, where the $N_a$ ensemble atoms are in the ground state, and state $\ket{\bf{r}}=\frac{1}{\sqrt{N_a}}\sum_{i=1}^{N_a}\ket{g_1}\ket{g_2}\cdots\ket{r_i}\cdots\ket{g_{N_a}}$, with a collective excitation. State $\ket{\bf{r}}$ is a superposition where one, and only one, of the atoms has been excited to the $\ket{r}$ state. In an ensemble, the atoms are indistinguishable and single photon absorption will always produce the same collective state. 

The ensemble can be thought of as a superatom with enhanced absorption properties. In principle, any atom with a transition resonant to the photon's frequency could be used as the quantum bomb. However, unless there is a strong coupling, the probability of absorption will be small and the scheme will not function as required. For that reason, we prefer using an ensemble, which is more delicate from the point of view of coherence time, but can provide a good probability of absorption when the bomb is supposed to be absorptive. As the ensemble is only an intermediate system, coherence time is not critical.

\begin{figure}[ht]
\centering
\includegraphics{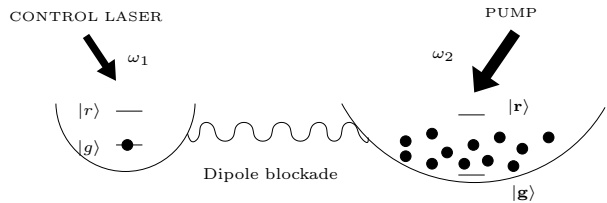}
\caption{Quantum bomb with Rydberg atoms: A single Rydberg atom controls the transmission properties of an atomic ensemble using dipole blockade.\label{Rydberg}} 
\end{figure}

Our bomb has two parts, an ensemble in the ground state $\ket{\bf{g}}$ and a control Rydberg atom in a superposition $\frac{\ket{g}+\ket{r}}{\sqrt{2}}$ (Figure \ref{Rydberg}). The photons will go through the ensemble. We consider two bomb states, $\0_B\equiv\ket{g}\ket{\bf g}$, in which the ensemble absorbs the photon with high probability, and $\1_B\equiv\ket{r}\ket{\bf g}$, in which the excited control atom interacts with the ensemble and raises the energy level of the excited states inhibiting photon absorption (there is a dipole blockade). 

Our scheme is similar to CNOT gate proposals for atom ensembles \cite{LFC01,MLW09}. We can recycle many techniques from previous quantum information systems based on Rydberg atoms \cite{SWM10}. However, instead of needing an exquisite control of the atoms, we shift most of the work into the quantum interrogation part. We only need to manipulate a single Rydberg atom and make it evolve from $\ket{g}$ into $\frac{\ket{g}+\ket{r}}{\sqrt{2}}$ (for instance with a controlled Rabi oscillation with a $\frac{\pi}{2}$ pulse). The bomb explosion (the photon absorption) can be a two photon resonant process assisted with a strong laser pump. In a correct operation, absorption will have a negligible probability, but it must be an efficient process. We need to guarantee the possibility of photon loss. The explosion can be amplified to the classical level, if desired, using fluorescence detection to observe whether the $\ket{r}$ state is occupied or not \cite{SPV04,SPV05}. This last step is optional. Quantum interrogation also works if the photons are lost to the environment when the bomb is present \cite{Whi98}. This includes absorption and a later spontaneous decay or any other incoherent process.

The ensemble states $\ket{\bf g}$ and $\ket{\bf r}$ are also valid absorbing and transparent bomb states. We have chosen to place the control in a single atom to avoid problems with the decay of the excited $\ket{\bf r}$ state. The single atom Rydberg state $\ket{r}$ is more stable and is separated from the passing photons, reducing the stimulated emission problems that could occur in the highly coupled photon-ensemble interaction. 

\begin{figure}[ht]
\centering
\includegraphics{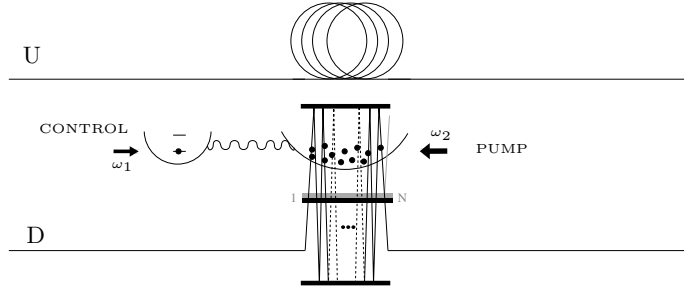}
\caption{Light-matter CZ gate with quantum interrogation: The state of the Rydberg bomb controls the phase difference between the two photon paths.\label{RydbergCZ}} 
\end{figure}

One possible implementation for the CZ gate between the ensemble and the photon is shown in Figure \ref{RydbergCZ}. This setup gives the evolution described in Equations (\ref{EqCZ1}) and (\ref{EqCZ2}).

The lower path of the photon crosses the same beamsplitter $N$ times and is directed with the help of two long mirrors. We place the ensemble, for instance a vapour cell, in the upper half of the interferometer. The upper photon, which does not enter the quantum interrogation part, is diverted through a path of the same length.

\section{The two photon gate}
The CZ gate of the previous section can be used to design a CZ gate between the two photons. The ensemble acts as an ancillary system which can be discarded after the operation. We start with two photons, a control photon in any superposition $\alpha_C\ket{1}_U^C\ket{0}_D^C+\beta_C\ket{0}_U^C\ket{1}_D^C$ and a target photon in state $\alpha_T\ket{1}_U^T\ket{0}_D^T+\beta_T\ket{0}_U^T\ket{1}_D^T$. 

The photonic CZ gate has six stages:
\begin{enumerate}
\item We initialize the bomb state to $\frac{\ket{g}\ket{\bf g}+\ket{r}\ket{\bf g}}{\sqrt{2}}$ acting on the control atom. If it is originally in the $\ket{g}$ state, we can apply a $\frac{\pi}{2}$ pulse.

\item We direct the control photon into the ensemble inside the quantum interrogation setup. The joint state of the bomb and the control photon goes from
\begin{equation}
\frac{\ket{g}\ket{\bf g}+\ket{r}\ket{\bf g}}{\sqrt{2}}\left(\alpha_C\ket{1}_U^C\ket{0}_D^C+\beta_C\ket{0}_U^C\ket{1}_D^C\right)
\end{equation}
to
\begin{equation}
\alpha_C\frac{\ket{g}\ket{\bf g}+\ket{r}\ket{\bf g}}{\sqrt{2}}\ket{1}_U^C\ket{0}_D^C+\beta_C\frac{\ket{g}\ket{\bf g}-\ket{r}\ket{\bf g}}{\sqrt{2}}\ket{0}_U^C\ket{1}_D^C.
\end{equation}

\item We apply a $\frac{\pi}{2}$ pulse on the control atom to get the state
\begin{equation}
\alpha_C\ket{g}\ket{\bf g}\ket{1}_U^C\ket{0}_D^C+\beta_C\ket{r}\ket{\bf g}\ket{0}_U^C\ket{1}_D^C.
\end{equation}
The state of the control photon is now associated to the transparency of the bomb. Control logical $\0_L^C$ becomes associated to an absorbing bomb, logical $\1_L^C$ to a transparent bomb.

\item We direct the target photon to the quantum interrogation setup with the ensemble. We have the evolution
\begin{eqnarray}
\alpha_C\ket{g}\ket{\bf g}\ket{1}_U^C\ket{0}_D^C\left(\alpha_T\ket{1}_U^T\ket{0}_D^T+\beta_T\ket{0}_U^T\ket{1}_D^T\right)+&&\nonumber\\
\beta_C\ket{r}\ket{\bf g}\ket{0}_U^C\ket{1}_D^C\left(\alpha_T\ket{1}_U^T\ket{0}_D^T+\beta_T\ket{0}_U^T\ket{1}_D^T\right)&&\rightarrow\\
\alpha_C\ket{g}\ket{\bf g}\ket{1}_U^C\ket{0}_D^C\left(\alpha_T\ket{1}_U^T\ket{0}_D^T+\beta_T\ket{0}_U^T\ket{1}_D^T\right)+&&\nonumber\\
\beta_C\ket{r}\ket{\bf g}\ket{0}_U^C\ket{1}_D^C\left(\alpha_T\ket{1}_U^T\ket{0}_D^T-\beta_T\ket{0}_U^T\ket{1}_D^T\right).&&
\end{eqnarray}

There is still a correlation to the state of the Rydberg atom which must be erased to avoid unwanted state changes.

\item We apply a $\frac{\pi}{2}$ pulse on the control atom. Now we have
\begin{eqnarray}
\alpha_C\frac{\ket{g}\ket{\bf g}+\ket{r}\ket{\bf g}}{\sqrt{2}}\ket{1}_U^C\ket{0}_D^C\left(\alpha_T\ket{1}_U^T\ket{0}_D^T+\beta_T\ket{0}_U^T\ket{1}_D^T\right)+&&\nonumber\\
\beta_C\frac{\ket{g}\ket{\bf g}-\ket{r}\ket{\bf g}}{\sqrt{2}}\ket{0}_U^C\ket{1}_D^C\left(\alpha_T\ket{1}_U^T\ket{0}_D^T-\beta_T\ket{0}_U^T\ket{1}_D^T\right).&&
\end{eqnarray}

\item We measure the state of the control Rydberg atom. If it is found in state $\ket{g}$, we already have the desired state 
\begin{equation}
\alpha_C\ket{1}_U^C\ket{0}_D^C\left(\alpha_T\ket{1}_U^T\ket{0}_D^T+\beta_T\ket{0}_U^T\ket{1}_D^T\right)+\beta_C\ket{0}_U^C\ket{1}_D^C\left(\alpha_T\ket{1}_U^T\ket{0}_D^T-\beta_T\ket{0}_U^T\ket{1}_D^T\right).
\end{equation}

If $\ket{r}$ is found, there is a sign shift in the terms associated to control logical $\1_L^C$. This can be corrected with a $\pi$ phase shifter in the lower path of the control photon. The phase shifter can be controlled with a driver connected to the measurement results. 

\end{enumerate}

After the sixth step, the photons are entangled and independent from the Rydberg atoms. We already have a gate which enables universal quantum computation with photons. 

\section{Experimental challenges}
The advantage of using a quantum interrogation scheme is that the degree of control on the ensemble can be limited. Absorption only happens in the case of failure and being able to keep the coherence of the $\ket{\bf r}$ state is irrelevant. The number of operations on the atomic part is likewise reduced. We only need to provide a suitable $\frac{\pi}{2}$ pulse for the control atom and a measurement system. The quantum interrogation approach shifts the problems to other areas. The most important obstacles are obtaining a good blockade, imperfect absorption and optical losses. 

We need a system showing a sufficiently strong dipole blockade so that the $\1_B$ bomb state is truly transparent. There has been enough experimental development in that area to guarantee at least a high transparency \cite{SWM10}. Small errors are not critical. We can monitor the system to find any residual absorption (bomb explosion). Protocol failures are heralded by a populated $\ket{\bf r}$ state, which can be observed. As long as the total probability of failure is kept small, we can just repeat the gate until we succeed. 

Imperfect absorption is another potential problem. We have described a system where the probability of photon absorption is 1 for the $\0_B$ bomb state. The real probability of absorption is limited by the photon-ensemble coupling. However, less than perfect absorption can be tolerated. There are different analyses of quantum interrogation of partially absorbing objects \cite{Jan99,KSS00,Azu06,GC05b}. They all predict the same asymptotic behaviour as in the case with a perfect absorber at the cost of increasing the number of cycles $N$. Furthermore, if the alternative bomb state is perfectly transparent, we can simply readjust the beamsplitters to recover perfect operation \cite{MM01}.

Losses pose a bigger challenge. Even for small optical losses, the number of stages before serious degradation is greatly reduced \cite{KWM99,Rud00}. A smaller $N$ can affect the final probability of success. A practical implementation of the proposed gate must achieve a delicate balance between losses and good absorption. Optically dense media with a higher interaction can increase the probability of loosing a photon when it must be preserved. 

\section{Outlook}
We have proposed an entangling gate for photons which, in combination with linear optics quantum gates, would allow to perform any quantum algorithm. The gate is based on the quantum interrogation of an atomic ensemble which can be in a superposition of being transparent and opaque to the photons. The transparency is turned on and off by manipulating a single Rydberg atom. If the atom is excited, it blocks absorption in the ensemble (the bomb is defused).  The interaction between the photons and the ensemble is based on the possibility of absorption, even though it never materializes during the correct operation of the gate. 

The main obstacle for a scalable quantum computer with such a counterfactual gate is achieving an optical system in which loss and absorption effects are properly balanced. While there are many additional technical difficulties, systems similar to the one proposed have been experimentally used for counterfactual interaction-free protocols with good results \cite{WIY11}. We believe that, even though the scheme can be refined, the implementation of, at least, the CZ gate between a single photon and the Rydberg atom system is feasible with the current technology.

\section*{Acknowledgements}
This research has been funded by Junta de Castilla y Le\'on project VA342B11-2 and MICINN TEC2010-21303-C04-04.

\newcommand{\noopsort}[1]{} \newcommand{\printfirst}[2]{#1}
  \newcommand{\singleletter}[1]{#1} \newcommand{\switchargs}[2]{#2#1}

\end{document}